\documentclass[twocolumn,secnumarabic,amssymb, nobibnotes, aps, prd]{revtex4-1}
\usepackage{graphicx}
\usepackage{subfloat}
\usepackage{subcaption}
\usepackage{xcolor}
\colorlet{blue}{black}

\begin{document}

\title{Intermediate scattering function and quantum recoil in non-Markovian quantum diffusion}

\author{Peter S.M. Townsend}
\email{psmt2@cam.ac.uk}
\affiliation{Surfaces, Microstructure and Fracture Group, Department of Physics, Cavendish Laboratory, J.J. Thomson Avenue, Cambridge, UK}

\author{Alex W. Chin}
\affiliation{Centre National de la Recherche Scientifique, Institute des Nanosciences de Paris, Sorbonne Universit\'{e}, Paris, France}

\date{\today}

\begin{abstract}
Exact expressions are derived for the intermediate scattering function (ISF) of a quantum particle diffusing in a harmonic potential and linearly coupled to a harmonic bath. The results are valid for arbitrary strength and spectral density of the coupling. The general, exact non-Markovian result is expressed in terms of the classical velocity autocorrelation function, which represents an accumulated phase during a scattering event. The imaginary part of the exponent of the ISF is proportional to the accumulated phase, which is an antisymmetric function of the correlation time $t$. The expressions extend previous results given in the quantum Langevin framework where the classical response of the bath was taken as Markovian. For a special case of non-Markovian friction, where the friction kernel decays exponentially in time rather than instantaneously, we provide exact results relating to unconfined quantum diffusion, and identify general features that allow insight to be exported to more complex examples. The accumulated phase as a function of the $t$ has a universal gradient at the origin, depending only on the mass of the diffusing system particle. At large $t$ the accumulated phase reaches a constant limit that depends only on the classical diffusion coefficient and is therefore independent of the detailed memory properties of the friction kernel. Non-Markovian properties of the friction kernel are encoded in the details of how the accumulated phase switches from its $t\rightarrow -\infty$ to its $t\rightarrow +\infty$ limit, subject to the constraint of the universal gradient. When memory effects are significant, the transition from one limit to the other becomes non-monotonic, owing to oscillations in the classical velocity autocorrelation. The result is interpreted in terms of a solvent caging effect, in which slowly fluctuating bath modes create transient wells for the system particle.
\end{abstract}

\maketitle

\section{Introduction and motivation}
\label{sec:intro}

The interaction of a quantum system with a thermal environment is a rich topic that arises naturally in many fields of physics, including quantum optics \cite{Raimond2001RevModPhys,Kok2007RevModPhys}, chemical physics \cite{Kondo1984PhysicaB}, nuclear physics \cite{Bertlmann2006PRA} and macroscopic quantum coherence \cite{Friedman2000Nature, Leggett1987RevModPhys}. The effect of the environment has a crucial bearing on foundational areas such as quantum measurement theory \cite{Walls1985PRD}, and on applications with the potential for enormous impact such as quantum computing \cite{Ladd2010Nature}. In the latter case, central questions include the precision to which coherence can be maintained in the presence of an environment, when a low-dimensional quantum system evolves from an initial superposition with a well-defined phase relationship. Loss of coherence, as well as population transfer, can be addressed on a consistent mathematical footing by considering the evolution of the qubit's reduced density matrix $\rho_{S}(t)$ in which the environment degrees of freedom are traced over \cite{Breuer2002OQS}. A range of techniques for time evolving $\rho_{S}(t)$ have been developed using, for example, projection operator techniques \cite{Zwanzig1960JCP}, stochastic wavefunction evolution \cite{Dalibard1992PRL} path integral methods \cite{Feynman1963AnnPhys}, and many-body wavefunction techniques that simulate the extended global system \cite{Schollwock2005RevModPhys}. There is no completely general, efficient method and so simplifying assumptions are required in different cases, for example weak coupling or the absence of memory effects in the thermal bath. Memory effects, when the effect of the bath cannot be treated in a Markovian approximation, are a wide area of topical interest in the field \cite{Breuer2015RevModPhys,DeVega2017RevModPhys}. It is widely recognised that exactly solvable large quantum systems, such as globally harmonic systems, are valuable for investigating the effect of arbitrarily strong non-Markovian quantum damping \cite{Breuer2002OQS}. In the present work we consider memory effects in a globally harmonic model that has been utilised extensively as a model for damped quantum oscillations including the unconfined limit of dissipative quantum diffusion \cite{Ford1965JMathPhys}.

In many contexts, such as chemical dynamics, the theoretical aim is to describe the dynamics of a system or particle in continuous contact with its environment, without being prepared in a special state to begin with. Then, the most convenient description of the open system dynamics is often not via the evolution of $\rho_{S}(t)$, but by equilibrium correlation functions $\langle A(t)B(0)\rangle$, the expectation of a product of operators evaluated at different times. Equilibrium correlation functions arise naturally \cite{VanHove1954PhysRev} in the description of experiments where the system dynamics are measured via a gentle scattering probe, for example in surface diffusion measurements with the helium-3 surface spin echo technique \cite{Jardine2009PCCP}. In surface diffusion, the strength and memory properties of the environment coupling play a central role in governing the rate and the detailed mechanism of dynamical processes, even within an entirely classical description and regardless of whether the diffusion is continuous or occurs by jumps \cite{Grote1980JCP, Hanggi1990RevModPhys, MiretArtes2005JPCM}. Surface diffusion in real physical systems takes place in a nonlinear potential energy landscape, and exact correlation functions are intractable. However, diffusion in either a flat or harmonic potential, coupled to a harmonic bath, can be described by a globally harmonic system and therefore exact thermal correlation functions can be derived both classically and quantum-mechanically because the global dynamics and thermodynamics are accessible. As with the problem of time-evolving $\rho_{S}$, correlation-function methods that can be used to treat nonlinear systems are generally restrictive in other ways, and establishing exact reference results for linear systems is therefore valuable for understanding the strengths and limitations of approximate methods. As an example, a formula has recently been proposed for calculating dynamical correlation functions for a particle in a periodic potential directly from the Bloch states of the uncoupled system and their lifetimes \cite{Firmino2014JPCL}. Exact results on harmonic systems can be used to explore the extent to which such methods can be pushed with respect to strong coupling and memory effects. 

Here we focus on the intermediate scattering function (ISF), the autocorrelation of the kinematic scattering amplitude $\exp(i\Delta\mathbf{K}\cdot\mathbf{x})$, which reflects the origins of the present work in the context of quasielastic atom-surface scattering \cite{Jardine2009PCCP}. A precise definition of the ISF will be given in Section \ref{sec:quantumISFDerivation} where the analytical results are derived. The significance of the ISF arises via a Born approximation for the inelastic scattering amplitude from dynamical scattering centres \cite{VanHove1954PhysRev}, in which the inelastic differential scattered intensity is proportional to the dynamical structure factor (DSF) of the ensemble of scattering centres. The ISF is the Fourier transform of the DSF into the time domain \cite{Lovesey1984Book}, and is measured approximately in the HeSE experiment where the Fourier transform is carried out physically \cite{Jardine2009PCCP}. The ISF is often the more convenient of the two scattering functions to work with, since closed analytical forms are available for a wide range of physical models including classical Langevin dynamics \cite{MiretArtes2005JPCM}. The short time behaviour of the ISF is sensitive to the nature of the coupling between each diffusing adsorbate and the substrate heat bath. For example, in the classical diffusion of an adsorbate subject to the Langevin equation, the ISF displays a regime switch between a Gaussian function describing ideal ballistic motion at short times, and an exponential decay describing continuous diffusion at long times. The crossover is compactly represented in the time domain, and the transition between the two regimes is governed by the velocity correlation time \cite{MiretArtes2005JPCM}. The classical result can be readily extended to cover the case of linear but non-Markovian dissipation \cite{Townsend2018JPCO}, and a key qualitative outcome is that the introduction of a finite memory time in the bath can strongly increase the amplitude of the ISF encompassed by ballistic-like behaviour, while leaving the long-term diffusion rate unaffected. Hence, the short-time behaviour of the classical ISF is sensitive to both the absolute strength of the coupling to the heat bath, and also the frequency dependence of the coupling. Later on we show that the same is true for the quantum mechanical ISF.

While the classical ISF is a real, symmetric function of the correlation time, the quantum ISF is complex. The origin of the complexity can be viewed as originating from the necessary asymmetry in the Fourier domain, a condition known as detailed balance imposed by the Boltzmann distribution \cite{Lovesey1984Book}. Equivalently, the origin of the imaginary contribution to the exponent of the ISF can be attributed to the position operator $x(t)$ of the scattering centre failing to commute with its original self $x(0)$ as it evolves in time via the operator equations of motion \cite{Martinez-Casado2008JCP, Martinez-Casado2010ChemPhys}. For a particle completely decoupled from its thermal bath and therefore carrying out ballistic motion, the result is a pure frequency-domain shift of the classical DSF, where the size of the shift is given by the dispersion relation of the scattering centre. Therefore in the time domain, the quantum ISF consists of the classical ISF multiplied by a non-decaying complex exponential in time. In the dissipative case, in which classically the particle undergoes a ballistic-diffusive transition, the imaginary part of the exponent of the ISF does not oscillate forever but is a damped, antisymmetric function of time whose limit as $t\rightarrow \infty$ is nonzero. The imaginary part is proportional to $\hbar$ and therefore describes a quantum effect, and its existence is known as quantum recoil \cite{Martinez-Casado2008JCP}. The functional form of the imaginary part of the exponent has been derived from a quantum Langevin description, both heuristically and in a linear response framework \cite{Martinez-Casado2008JCP,Martinez-Casado2010ChemPhys}, assuming that the classical fluctuation and dissipation are Markovian. The value of the present work in relation to those previous studies will be to give a concise expression for the imaginary part of the ISF exponent in terms of the classical velocity autocorrelation function, and evaluate the function for an example of non-Markovian linear dissipation.

The linear coupling model of surface diffusion is a well explored model system and has been investigated in some detail using projection operator methods \cite{Breuer2001AnnPhys} and functional integral approaches \cite{Caldeira1983PhysicaA,Grabert1988PhysRep}. Additionally, atom-scattering line shapes have been derived for scattering from surface phonons and harmonically bound adsorbates using fundamentally the same model \cite{MartinezCasado2010JPCM}. However, to our knowledge the precise analytical connection between linear correlation functions and quantum recoil in the ISF, for non-Markovian coupling to the bath, has not been fully elucidated, and that is the purpose of the present work. In Section \ref{sec:quantumISFDerivation}, exact expressions are given for the real and imaginary parts of the exponent of the ISF in terms of classical correlation functions. The results are valid for any globally harmonic system, and could therefore apply to damped vibrations as well as dissipative diffusion. The imaginary part is illustrated for the special case of exponential memory friction where memory effects are described by a single parameter and the classical velocity autocorrelation is straightforwardly accessible.

\section{Derivation of the quantum ISF}
\label{sec:quantumISFDerivation}

Consider the model Hamiltonian $H$ in which a particle of mass $m$, harmonically bound in a one-dimensional potential of natural oscillation frequency $\omega_{0}$, is linearly coupled to a harmonic bath as follows:
\begin{equation}
\label{eqn:calLegHam}
H=\frac{p^{2}}{2m} + \frac{1}{2}m\omega_{0}^{2}+\sum_{\alpha} \Bigg[ \frac{p_{\alpha}^{2}}{2m_{\alpha}}+\frac{1}{2}m_{\alpha}\omega_{\alpha}^{2}\Big(x_{\alpha}-\frac{c_{\alpha}x}{m_{\alpha}\omega_{\alpha}^{2}}\Big)^{2} \Bigg]\,\textrm{.}
\end{equation}
$x$ and $p$ are the position and co-ordinate operators of the particle considered as our open system. $p_{\alpha}$ and $x_{\alpha}$ are the position and momentum operators of bath degrees of freedom, which are harmonic oscillators of frequency $\omega_{\alpha}$ and mass $m_{\alpha}$, coupled to our system particle with coupling constants $c_{\alpha}$.

The quantum ISF for our system particle is defined as
\begin{equation}
\label{eqn:isfDef}
I(\Delta K,t)=\frac{1}{Z}\mathrm{tr}\Big[ e^{i\Delta K x(t)} e^{-i\Delta K x(0)}e^{-\beta H}\Big] \textrm{,}
\end{equation}
where $Z$ is the partition function of the global model, $\Delta K$ is a parameter called the momentum transfer, $\beta$ is the inverse temperature $(k_{B}T)^{-1}$, and any complete basis can be chosen for the trace. \textcolor{blue}{The definition is made within the Heisenberg picture of operator evolution, in which $x(t)$ is the time evolution of $x(0)$. Throughout the present article, operators without an explicit time argument have an implicit time argument of zero.} The connection between the definition here and the dynamic structure factor and hence scattering intensity in the Born approximation, can be established by performing the trace in the basis of global energy eigenstates.

A real scattering experiment would normally involve scattering of a beam of particles from an ensemble of scattering centres, say an ensemble of atoms adsorbed on a surface. Here we are assuming that no scattering is induced directly by the environment modes (such as phononic or electronic excitations). Additionally, if the probe particles scatter coherently from the adsorbates in the ensemble, then we are assuming that there are no explicit correlations between the dynamics of the different particles in the ensemble, which is a reasonable qualitative assumption as long as the ensemble has a low density. However, our purpose here is to derive an exact result on a model system rather than account for the additional factors that would affect the results of approximate experimental realisations.

If $H$ is considered as a classical Hamiltonian, then the classical dynamics of the system particle are given by the generalised Langevin equation (GLE) \cite{Cortes1985JCP,Weiss2012QDS},
\begin{equation}
\label{eqn:CalLegGLE}
m\ddot{x}(t)=-m\omega_{0}x(t)-\int_{0}^{t} m\gamma(t-t')\dot{x}(t')dt'\, + F(t) \, \textrm{,}
\end{equation}
with the friction kernel given by
\begin{equation}
\label{eqn:CalLegKernel}
\gamma(t)=\theta(t)\sum_{\alpha}\frac{c_{\alpha}^{2}}{m_{\alpha}\omega_{\alpha}^{2}}\,\cos(\omega_{\alpha}t)
\end{equation}
and where $F(t)$ is a normally distributed, zero-mean random force that satisfies $\langle F(t)F(0) \rangle = m k_{B}T\gamma(t)$, a classical fluctuation-dissipation relation where $\langle \rangle$ indicates an ensemble average over initial states of the bath with the position of the system particle taken into account in the averaging process \cite{Cortes1985JCP}. The GLE can be used to derive classical correlation functions such as the velocity autocorrelation function (VACF) $\psi(t)=\langle v(t)v(0)\rangle$, which can be readily expressed as a Laplace transform, and expressed analytically in the time domain whenever the Laplace transform is analytically invertible. For example, when the friction kernel $\gamma(t)$ is an exponentially decaying function of time, the resulting Laplace space form of the VACF can be straightforwardly inverted to give a biexponential function in time \cite{Berne1966JCP} which generalises the well-known mono-exponential form $\psi(t)=\langle v^{2}\rangle \exp(-\gamma t)$ derived from the Langevin equation \cite{MiretArtes2005JPCM}.

In the classical situation just described, the effect of the bath mode masses $m_{\alpha}$ on the dynamical properties on the system particle is entirely captured in the expansion (\ref{eqn:CalLegKernel}) where the masses always enter in the combination $c_{\alpha}^{2}/m_{\alpha}$. Therefore, $c_{\alpha}$ can always be traded against $m_{\alpha}$ to make the bath mode masses equal to the system particle mass ($m_{\alpha}=m \forall \alpha$) with no loss of generality, as long as we are interested only in correlation functions involving the system particle alone. The same outcome can be seen where, for example, an explicit transformation to mass-weighted co-ordinates has been used to address memory effects in classical barrier crossing \cite{Pollak1986JCP}. The operator-valued generalisation of the classical GLE (\ref{eqn:CalLegGLE}) is the quantum Langevin equation (QLE) for the system particle's position operator which reads the same as the GLE but for operator-valued $x(t)$ and $F(t)$, where quantum effects enter into the fluctuation-dissipation relations that apply to $F$ \cite{Ford1987JStatPhys}. In the present notation \cite{Weiss2012QDS}, the statistical dynamics of the random force (still with zero mean) are given in terms of the commutator $[A,B]=AB-BA$ and the anti-commutator $\{A,B\}=AB+BA$ by
\begin{equation}
\label{eqn:randomForceAnticomm}
\langle \, [F(t),F(0)] \, \rangle = -i\hbar \sum_{\alpha} \frac{c_{\alpha}^{2}}{m_{\alpha}\omega_{\alpha}}\,\sin(\omega_{\alpha}t)\textrm{;}
\end{equation}
\begin{equation}
\label{eqn:randomForceComm}
\langle\, \{ F(t),F(0) \} \, \rangle = \hbar \sum_{\alpha} \frac{c_{\alpha}^{2}}{m_{\alpha}\omega_{\alpha}}\,\coth\Big(\frac{1}{2}\beta\hbar\omega_{\alpha}\Big)\,\cos(\omega_{\alpha}t)\,\textrm{,}
\end{equation}
where in the quantum mechanical case as the classical case, the averaging $\langle \rangle$ is performed over a bath equilibrated with the initial system particle position \cite{Cortes1985JCP,Weiss2012QDS}. The numbers $\{c_{\alpha}\}$ and $\{m_{\alpha}\}$, characterising the bath, still appear only in the combination $c_{\alpha}^{2}/m_{\alpha}$. Therefore, just as in the classical case, being able to vary $c_{\alpha}$ and $m_{\alpha}$ independently gives no more flexibility than fixing $m_{\alpha}=m\forall\alpha$ and varying $c_{\alpha}$, in terms of the effect on system particle correlation functions. The purpose of writing out the QLE explicitly in the present work was to emphasise the amount of generality retained even when $m_{\alpha}=m\forall\alpha$; the derivation of the ISF will proceed shortly through a different representation of the system-bath coupling based on the global normal modes.

As a separate convenient ingredient for the derivation, we quote a re-exponentiation result for the ISF of a particle in a harmonic potential but not coupled to a bath. Namely, the ISF associated with the uncoupled Hamiltonian
\begin{equation}
H_{\Omega}=\frac{p^{2}}{2m}+\frac{1}{2}m\Omega^{2}x^{2}
\end{equation}
is given by \cite{Gunn1984ZPB}:
\begin{equation}
\label{eqn:basicExponentiationFormula}
I_{\Omega}(\Delta K,t)=\exp\Bigg\{\frac{1}{2}\Delta K^{2}\Big[X_{\Omega}(t)+iY_{\Omega}(t)\Big]\Bigg\}\,\textrm{,}
\end{equation}
where
\begin{equation}
X_{\Omega}(t)= \frac{1}{m\Omega}\Big[\,\cos(\Omega t)-1\Big]\hbar\,\coth\Big(\frac{1}{2}\beta\hbar\Omega \Big) \,\textrm{;}
\end{equation}
\textcolor{blue}{
\begin{equation}
i\hbar Y_{\Omega}(t)= i\hbar\frac{1}{m\Omega}\,\sin(\Omega t) \,\textrm{.}
\end{equation}
}
Taking the limit $\Omega\rightarrow 0$ returns the quantum ballistic ISF that can be obtained directly using, for example, the Baker-Hausdorff theorem \cite{Martinez-Casado2008JCP}, which demonstrates that it is safe to treat a free particle as the $\omega_{0}\rightarrow 0$ limit of a quantum oscillator in the present context. Next we consider the ISF (\ref{eqn:isfDef}) when the scattering centre is coupled to the environmental oscillators.

\textcolor{blue}{To compute the ISF for the open system, we put the system and bath co-ordinates are on equal footing by performing a normal modes transformation, a simultaneous orthogonal transformation of the co-ordinates and momenta of the global model such that the Hamiltonian as a function of the new operators represents a collection of uncoupled oscillators. An application of normal modes transformations to study classical barrier crossing has been mentioned already \cite{Pollak1986JCP}; in the quantum mechanical case the transformation is also known as a Bogoliubov-Valatin transformation \cite{TaylorHeinonen2004}, when considered as a transformation of creation and annihilation operators. In general, the purpose of such a transformation is to find the good quantum numbers of collective excitations. Phonons in a harmonic solid provide one familiar example, but the technique also finds broad application across condensed matter, in systems that can be described via a variable number of collective excitations, including superfluidity and magnetism \cite{Altland2007CMFT}.} 

\textcolor{blue}{The normal modes transformation brings the Hamiltonian into the form}
\begin{equation}
\label{eqn:normalModesHamiltonian}
H=\sum_{k} \Big( \frac{q_{k}^{2}}{2m} + \frac{1}{2}m_{k}\Omega_{k}^{2}y_{k}^{2} \Big) \,\textrm{,}
\end{equation}
\textcolor{blue}{where $y_{k}$ are the quantum operators representing normal co-ordinates, $q_{k}$ are the corresponding canonical momenta, $m$ is still the particle mass, and $\{\Omega_{k}\}$ are the frequencies of the oscillators that have been decoupled by the transformation. The normal co-ordinates $y_{k}$ and corresponding momenta $q_{k}$ satisfy the canonical commutation relations as long as the original $x$, $p$ and $x_{\alpha}$, $p_{\alpha}$ did so, since the normal modes transformation is orthogonal and therefore canonical. Further details of the operator transformation are given in the Appendix. If there are $N$ bath modes, that is $\alpha$ runs from $1$ to $N$, then there are $N+1$ values of the index $k$. The system and bath modes are not treated separately by the $k$ index, and so when we compute the ISF shortly, there will be no separate summation over bath modes and system states, only a single summation over the normal modes of the global system.}
	
\textcolor{blue}{The frequency sets $\{\Omega_{k}\}$ and $(\omega_{0},\{\omega_{\alpha}\}$, and the corresponding coefficient sets $\{d_{k}\}$ and $\{c_{\alpha}\}$ are related via the solution of an eigenvalue problem, and there is no general expression for a specific element of one set in terms of the elements of the other. However, the fact that the Hamiltonian can be expressed in the form (\ref{eqn:normalModesHamiltonian}) allows formally exact expressions for the ISF to be found, which can in turn be related back to the original parameters of the coupled-modes Hamiltonian as we will see shortly. Therefore $\{d_{k}\}$ and $\{\Omega_{k}\}$ never need to be known explicitly if it is not convenient to compute them. As part of the definition of the model problem, the operator $x$ always represents the system co-ordinate, regardless of the values all other parameters. Aside from starting with the inherently approximate model of Equation \ref{eqn:calLegHam}, no further approximations are made concerning the system and bath frequencies, or the overall strength of coupling to the bath.}

We now consider one row of the normal modes transformation, namely the expression for $x$ in terms of the global normal mode co-ordinates, in terms of unknown coefficients $d_{k}$:
\begin{equation}
\label{eqn:normalModesFirstRow}
x=\sum_{k}d_{k}y_{k}\,\textrm{.}
\end{equation} 
It follows from the separable form (\ref{eqn:normalModesHamiltonian}) of the Hamiltonian that the ISF is the product of terms like that of Equation \ref{eqn:basicExponentiationFormula}, \textcolor{blue}{which we now demonstrate.}
\color{blue}
For convenience we write the separable Hamiltonian as the sum of commuting parts $H_{k}$,
\begin{equation}
H=\sum_{k} H_{k}\,\textrm{,}
\end{equation}
where
\begin{equation}
H_{k}=\frac{q_{k}^{2}}{2m} + \frac{1}{2}m_{k}\Omega_{k}^{2}y_{k}^{2}\,\textrm{.}
\end{equation}
We substitute the linear combination (\ref{eqn:normalModesFirstRow}) into the definition (\ref{eqn:isfDef}) of the ISF, and take the trace in the basis of eigenstates of the normal mode co-ordinate operators. If we define $A$ as the operator whose trace gives the numerator of the ISF, namely $A=e^{i\Delta K x(t)}e^{i\Delta K x(0)}e^{-\beta H}$, with the explicit expansion in normal co-ordinate operators
\begin{equation}
A=e^{i\Delta K \sum_{k}d_{k}y_{k}(t)} e^{-i\Delta K \sum_{k}d_{k}y_{k}(0)} e^{-\beta \sum_{k}H_{k}}\,\textrm{,}
\end{equation}
then
\begin{eqnarray}
I(\Delta K,t)=&\frac{1}{Z}\int d\mathbf{y} \langle \mathbf{y}| A |\mathbf{y}\rangle \, \textrm{,}
\end{eqnarray}
where $\int d\mathbf{y}$ stands for $\int dy_{1} dy_{2} \cdots dy_{N+1}$, and $|\mathbf{y}\rangle$ stands for $|y_{1}\rangle |y_{2}\rangle\cdots |y_{N+1}\rangle$. By construction, the operators associated with different normal modes commute, i.e. if $k\neq l$ then $[H_{k},H_{l}]=[y_{k},y_{l}]=[q_{k},q_{l}]=0$. Therefore the exponential operators can be arranged as a product over $\{k\}$, which holds for all times $t$ since the time evolution of the normal co-ordinate operators does not mix the different $k$.

The trace itself therefore also reduces to a product, where if we define operators
\begin{equation}
O_{k}(t)=e^{i\Delta K d_{k}y_{k}(t)} e^{-i\Delta K d_{k}y_{k}(0)}e^{-\beta \sum_{k}H_{k}}\,\textrm{,}
\end{equation}
then
\begin{equation}
I(\Delta K,t)=\frac{1}{Z} \prod_{k} \int dy_{k} \langle y_{k}|O_{k}(t) |y_{k}\rangle \, \textrm{.}
\end{equation}
We emphasise that the trace is not performed separately over bath and system degrees of freedom, which are mixed by the normal modes transformation.

By writing the partition function similarly as a product over $k$, the result can be written in terms of the one-mode ISF of Equation \ref{eqn:basicExponentiationFormula} as
\begin{equation}
I(\Delta K,t)=\prod_{k} I_{\Omega_{k}}(d_{k}\Delta K,t)\,\textrm{,}
\end{equation}
where each coupling coefficient $d_{k}$ is accounted for efficiently by noting that it appears exclusively in the combination $d_{k}\Delta K$.

\color{black}
The result can be conveniently written as
\begin{equation}
\label{eqn:realImExp}
I(\Delta K,t)=\exp\Bigg\{\frac{1}{2}\Delta K^{2}\Big[X(t)+iY(t)\Big]\Bigg\}\,\textrm{,}
\end{equation}
where
\begin{equation}
X(t)= \sum_{k}\frac{d_{k}^{2}}{m_{k}\Omega_{k}}\Big[\,\cos(\Omega_{k} t)-1\Big]\hbar\,\coth\Big(\frac{1}{2}\beta\hbar\Omega_{k}\Big) \,\textrm{;}
\end{equation}
\begin{equation}
i\hbar Y(t)= i\hbar\sum_{k}\frac{d_{k}^{2}}{m_{k}\Omega_{k}}\,\sin(\Omega_{k}t) \,\textrm{.}
\end{equation}

\textcolor{blue}{Since $\Delta K$ appears in the exponent of the one-mode formula (\ref{eqn:basicExponentiationFormula}) as $\Delta K^{2}$, the $k^{\mathrm{th}}$ contribution to the exponent in the multi-mode result is weighted by $d_{k}^{2}$.} The $d_{k}$ coefficients are as yet unspecified, but the real and imaginary parts of the exponent can be written entirely in terms of classical correlation functions of the system, which in turn depend on the classical friction kernel $\gamma(t)$. The friction kernel (\ref{eqn:CalLegKernel}) is given directly in terms of the original specification of the coupling constants $c_{\alpha}$. To draw the connection with classical correlation functions we first evaluate the classical VACF $\psi(t)$ in terms of the $d_{k}$. The classical velocity is given by the prevailing transformation into normal modes,
\begin{equation}
\dot{x}=\sum_{k} d_{k}\dot{y}_{k}\,\textrm{.}
\end{equation}
The time evolution of a normal mode is simply
\begin{equation}
y_{k}(t)=y_{k}(0)\,\cos(\Omega_{k}t)+\frac{q_{k}(0)}{m_{k}\Omega_{k}}\,\sin(\Omega_{k}t)\,\textrm{,}
\end{equation}
where $m_{k}=m \forall k$ because $m_{\alpha}=m \forall \alpha$ so that the normal modes transformation could be performed without any transformation of mode masses. The classical velocities therefore evolve according to:
\begin{equation}
\dot{y}_{k}(t)=\frac{q_{k}(0)}{m_{k}}\cos(\Omega_{k}t)-y_{k}(0)\Omega_{k}\cos(\Omega_{k}t)\,\textrm{.}
\end{equation}
Performing the thermal, classical phase space average over Boltzmann-distributed initial conditions $y_{k}(0)$ and $q_{k}(0)$ gives the VACF as
\begin{equation}
\label{eqn:normalModesVACF}
\psi(t)=\frac{k_{B}T}{m}\sum_{k} d_{k}^{2}\,\cos(\Omega_{k}t)\,\textrm{.}
\end{equation}
As a simple check on the consistency of the result, we recall that the coefficients $d_{k}$ form the row of an orthogonal matrix which effected the normal modes transformation, and therefore $\sum_{k} d_{k}^{2}=1$, which is consistent with the zero-time limit $\psi(0)=\langle v^{2} \rangle = k_{B}T/m$. It will be convenient now to define a normalised VACF,
\begin{equation}
\phi(t)=\frac{m}{k_{B}T}\psi(t)=\sum_{k} d_{k}^{2}\,\cos(\Omega_{k}t)\,\textrm{.}
\end{equation}
The imaginary part of the ISF exponent, $Y(t)$, can be written compactly in terms of $\phi(t)$ as:
\begin{equation}
\label{eqn:recoilFromVACF}
Y(t)=\frac{1}{m}\int_{0}^{t}\phi(t')dt'\,\textrm{.}
\end{equation}

Defining a new function $\psi_{Q}(t)$ as the classical VACF filtered by the function $\frac{1}{2}\beta\hbar\omega\,\coth(\frac{1}{2}\beta\hbar\omega)$ in the frequency domain, i.e.
\begin{equation}
\Psi_{Q}(t)=\frac{k_{B}T}{m}\sum_{k} d_{k}^{2}\, \frac{1}{2}\beta\hbar\Omega_{k}\,\coth\Big(\frac{1}{2}\hbar\beta\Omega_{k}\Big)\, \cos(\Omega_{k}t)\,\textrm{,}
\end{equation}
then the function $X(t)$ is given by an expression identical in form to the classical cumulant expansion \cite{MiretArtes2005JPCM} relating $I(\Delta K,t)$ and $\psi(t)$, namely:
\begin{equation}
\label{eqn:quantumCumulantExpansion}
-\frac{1}{2}X(t)=\int_{0}^{t}(t-t')\psi_{Q}(t')dt'\,\textrm{,}
\end{equation}
which is easily verified using the identity $\int_{0}^{t} \, dt' \, (t-t')\,\cos(\Omega t')=[1-\,\cos(\Omega t)]/\Omega^{2}$.

Therefore, the quantum ISF is not quite the product of the classical ISF and a quantum recoil factor, as the real part of the exponent has been filtered in a way that reflects the spectral density of the global normal modes, and quantum rather than classical occupation factors. However, the real part of the exponent can still be derived entirely from the classical VACF for the model system considered in the present work, by applying a Fourier filter. Alternatively, by evaluating the quantum mean square displacement (MSD) $\langle[x(t)-x(0)]^{2}\rangle$, it is readily shown that the result (\ref{eqn:quantumCumulantExpansion}) is equivalent to replacing the classical MSD in the classical cumulant expansion of the ISF \cite{MiretArtes2005JPCM} with the quantum MSD. For the remainder of the paper we will not consider $X(t)$ in further detail, but focus on the purely quantum-mechanical term $Y(t)$. 

\color{blue}
One of the key results of the present section is that the input parameters of the model required to evaluate Equation \ref{eqn:realImExp} can be specified in any of several forms. Any of the following inputs, in addition to the particle mass, would be sufficient to evaluate the model's quantum ISF:
\begin{enumerate}
	\item The parameters $\omega_{0}$, $\{\omega_{\alpha}\}$ and $\{c_{\alpha}\}$ of the model Hamiltonian expressed in the form of coupled oscillators. Assuming the bath modes form a continuum, the parameter set is conveniently expressed as the spectral density of the bath coupling, conventionally written as $J(\omega)=\frac{\pi}{2}\sum_{\alpha}\frac{c_{\alpha}^{2}}{m_{\alpha}\omega_{\alpha}}\delta(\omega-\omega_{\alpha})$ \cite{Weiss2012QDS}. The spectral density  can be derived for specialised model cases such as for a particle embedded in a harmonic chain \cite{Rubin1963PhysRev}, but could alternatively be specified as a phenomenological input without a rigorous derivation, chosen to represent the underlying physics or timescales of the environment.
	\item The parameters $\{d_{k}\}$ and $\{\Omega_{k}\}$ of the model Hamiltonian expressed in the form of decoupled oscillators.
 	\item The classical velocity autocorrelation $\psi(t)$, or friction kernel  $\gamma(t)$. Both time-dependent functions are readily related to the underlying parameters of the Hamiltonian via relations such as \ref{eqn:CalLegKernel} and \ref{eqn:normalModesVACF}. Further, $\psi(t)$ and $\gamma(t)$ are related to each other via a Laplace transform of the GLE (\ref{eqn:CalLegGLE}) \cite{Berne1966JCP}. Additionally, $\phi(t)$ and $\gamma(t)$ are also routinely computed from classical simulations of many-body anharmonic systems such as liquids \cite{Harp1970PRA}. Equations such as (\ref{eqn:recoilFromVACF}) applied to such simulation data would then represent a prediction of non-Markovian effects in quantum recoil within a Gaussian approximation to the anharmonic dynamics.
\end{enumerate}

\color{black}
\section{Quantum recoil subject to memory friction}
\label{sec:exponentialApplication}

To illustrate the new result concerning quantum recoil, the imaginary part of the ISF exponent can be calculated for a simple non-Markovian model. We consider an unconfined particle ($\omega_{0}=0$) undergoing quantum Brownian motion in which the classical friction kernel (\ref{eqn:CalLegKernel}) consists of an exponential decay in time. \textcolor{blue}{The unconfined, or flat-surface case, is chosen in order to simplify the analytical results as far as possible and to isolate the oscillatory features of the quantum recoil line shape that arise purely due to memory friction. However, the methods of Section \ref{sec:quantumISFDerivation} are still an indispensable part of the argument even when $\omega_{0}=0$, as the globally harmonic model Hamiltonian was a necessary step to derive the exact relationship between the classical velocity autocorrelation and the imaginary part of the quantum ISF. In other words we are representing non-Markovian dissipation by the globally harmonic model analysed in Section \ref{sec:quantumISFDerivation}.}

The \textcolor{blue}{exponential} kernel is described by two parameters $\gamma$ and $\omega_{c}$, as 
\begin{equation}
\label{eqn:exponentialKernel}
\gamma(t)=\theta(t)\gamma\omega_{c}e^{-\omega_{c}t}\,\textrm{.}
\end{equation}
As $\omega_{c}$ is varied, the total time integral of the friction kernel, or equivalently $\tilde{\gamma}(0)$ the kernel at zero frequency in the Fourier domain, is being kept constant. \textcolor{blue}{The classical VACF can then be derived from a Laplace transform of the classical GLE \cite{Berne1966JCP}. The result is}
\begin{equation}
\psi(t)=\frac{k_{B}T}{m} \Big(p_{1}e^{s_{1}|t|}+p_{2}e^{s_{2}|t|}\Big) \, \textrm{,}
\end{equation}
where $s_{1}$ and $s_{2}$ are the solutions of
\begin{equation}
\label{eqn:laplaceRoots}
s^{2}+\omega_{c}s+\gamma\omega_{c}=0 \,\textrm{.}
\end{equation}
and
\begin{equation}
p_{1}=\frac{(s_{1}+\omega_{c})}{s_{1}-s_{2}} \, \textrm{ ; } \, p_{2}=\frac{(s_{2}+\omega_{c})}{s_{2}-s_{1}} \,\textrm{.}
\end{equation}
The normalised VACF is
\begin{equation}
\phi(t)= p_{1}e^{s_{1}|t|}+p_{2}e^{s_{2}|t|} \, \textrm{,}
\end{equation}
and therefore the recoil function is given by:
\begin{equation}
\frac{t}{|t|}mY(t)=\Big( \frac{p_{1}}{s_{1}}e^{s_{1}|t|}+\frac{p_{2}}{s_{2}}e^{s_{2}|t|} \Big) - \Big( \frac{p_{1}}{s_{1}}+\frac{p_{2}}{s_{2}}\Big) \,\textrm{.}
\end{equation}
Using the properties of quadratic roots, the constant term in $Y(t)$ simplifies, giving
\begin{equation}
\label{eqn:recoilLongTimeLimit}
\frac{t}{|t|}mY(t)=\Big( \frac{p_{1}}{s_{1}}e^{s_{1}|t|}+\frac{p_{2}}{s_{2}}e^{s_{2}|t|} \Big) + \frac{1}{\gamma} \,\textrm{.}
\end{equation}
The limit of $Y(t)$ at large positive and negative times is therefore independent of $\omega_{c}$. There is a connection between the $\omega_{c}$-independence of the limits of $Y(t)$, and the $\omega_{c}$-independence of the classical diffusion coefficient $D$. The diffusion coefficient is given by \cite{Kubo1966RepProgPhys}
\begin{equation}
\label{eqn:kuboDiffCoeff}
D=\int_{0}^{\infty}\psi(t')\,dt'\,\textrm{,}
\end{equation}
but from the construction of $Y(t)$ as an accumulated phase governed by the velocity correlation, it follows that
\begin{equation}
D=k_{B}TY(\infty)\,\textrm{.}
\end{equation}
In other words, the classical diffusion coefficient governs the long-time limit of the recoil function. The result is as general as the relations (\ref{eqn:kuboDiffCoeff}) and (\ref{eqn:recoilFromVACF}), and therefore although it is neatly illustrated by the exponential kernel, the result is not dependent on any specific friction kernel.

\begin{figure}
\centering
\includegraphics[width=0.45\textwidth]{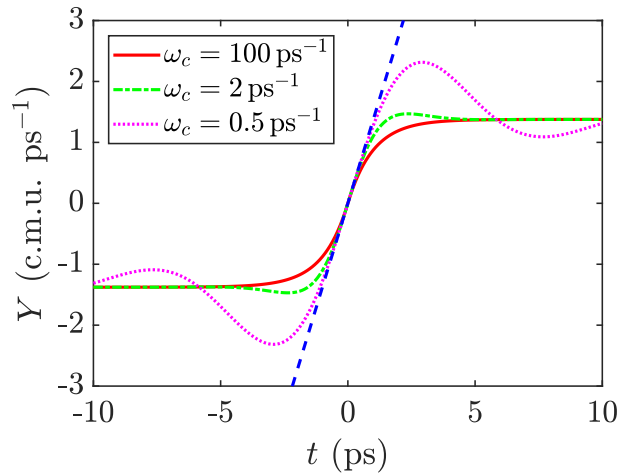}
\caption{Analytical forms of the recoil function $Y(t)$ for ballistic motion (blue dashed line) and different baths represented by the friction kernel (\ref{eqn:exponentialKernel}). One c.m.u. (approximately $0.1$ atomic mass units) is defined here as the mass unit consistent with a picosecond, {\AA}, meV system. The mass of the particle is $7.0$ atomic mass units; all else being equal, the size of the quantum recoil effect scales as $Y(t)\propto 1/m$. The friction coefficient $\gamma$ in $\gamma(t)=\theta(t)\gamma\omega_{c}e^{-\omega_{c}t}$ was taken as $\gamma=1.0\,$ps$^{-1}$, a ballpark figure applicable to the diffusion of adsorbates on metal surfaces. The key features of the curves with varying $\omega_{c}$ are a universal gradient at the origin, which matches the result for ballistic motion, and a limit depending only on $\gamma$ (not $\omega_{c}$) as $t\rightarrow\pm\infty$. Different values of $\omega_{c}$, shown in the legend, vary from $\omega_{c}\ll\gamma$ to $\omega_{c}\gg\gamma$. When $\omega_{c}$ is very large such that the friction is effectively Markovian, the recoil function transitions monotonically between the limits (red solid curve). When $\omega_{c}<4\gamma$ the VACF $\phi(t)$ acquires a cosine component and therefore oscillatory features are present in $Y(t)$ (dot-dashed green curve).}
\label{fig:quantumRecoil}
\end{figure}

Figure \ref{fig:quantumRecoil} shows the quantum recoil function $Y(t)$ for a particle of mass $7.0$ atomic mass units, subject to the exponential friction kernel (\ref{eqn:exponentialKernel}) with $\gamma=1.0\,$ps$^{-1}$ and different cutoff frequencies $\omega_{c}$, which include an essentially Markovian example ($\omega_{c}\gg \gamma$). Also shown is the $\gamma=0$ result, corresponding to ballistic motion of the system particle. The recoil function is always antisymmetric, due to its relationship to the Fourier transform of a real function $S(\Delta K,\omega)$. $Y(t)$ for ballistic motion is linear, with a gradient such that when the complex ISF (\ref{eqn:realImExp}) is reconstructed, its representation in the energy domain is simply the classical result but shifted by a recoil energy $E_{r}=\hbar^{2}\Delta K^{2}/2m$ \cite{Martinez-Casado2008JCP}. Comparing to the curves in the presence of the bath shows that the gradient at the origin is a universal property, independent of $\gamma$ or $\omega_{c}$. The universality can be understood on the basis that no matter how strong the coupling to a bath, on a short enough timescale the motion of a classical particle will always appear ballistic, with the bath imposing thermal initial conditions. The result therefore applies regardless of either the detailed form, or the absolute strength, of the friction kernel. Coupling to the bath leads to a finite, $\omega_{c}$-independent plateau value $Y(\pm\infty)=\pm 1/m\gamma$ as shown by (\ref{eqn:recoilLongTimeLimit}). When memory effects are unimportant the recoil function transitions smoothly between $Y(-\infty)$ and $Y(+\infty)$ over a transition time governed by $s_{1}$ and $s_{2}$ which tend to $\gamma$ when $\gamma\gg\omega_{c}$. However, when $\omega_{c}<4\gamma$, the decay rates $s_{1}$ and $s_{2}$ take complex values which gives rise to oscillations in $\phi(t)$, $\psi(t)$ and $Y(t)$.

\begin{figure}
\centering
\includegraphics[width=0.45\textwidth]{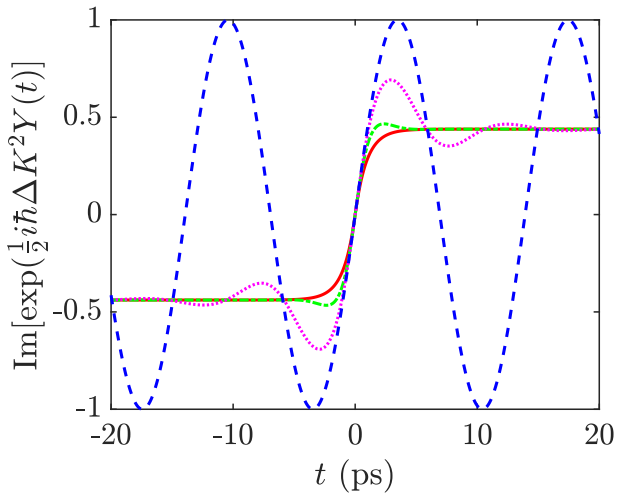}
\caption{\label{fig:quantumRecoilISFFactor} The recoil functions in Figure \ref{fig:quantumRecoil} have been exponentiated at $\Delta K=1.0\,${\AA}$^{-1}$ to give the complex factor $\exp(\frac{1}{2}i\hbar\Delta K^{2}Y(t)$ which appears in the ISF (\ref{eqn:realImExp}). The plot shows the imaginary part only. When the exponent is not too large, the shape of $Im[\exp(\frac{1}{2}i\hbar\Delta K^{2}Y(t))]$ is very similar to the shape of $Y(t)$ itself, due to the linear small-argument expansion of $\sin(\frac{1}{2}\hbar\Delta K^{2}Y(t))$. However, when the accumulated phase becomes very large, extended oscillations are seen, as shown by the result for ballistic motion (blue dashed curve).}
\end{figure}

Figure \ref{fig:quantumRecoilISFFactor} shows the imaginary part of the complex factor $\exp(\frac{1}{2}i\hbar\Delta K^{2}Y(t)$ in the ISF (\ref{eqn:realImExp}), derived from the recoil functions plotted in Figure \ref{fig:quantumRecoil}. In the damped examples, with the numerical parameters chosen, the plotted imaginary part has a similar form to the recoil function itself, since a small-argument approximation applies $\sin(\frac{1}{2}\hbar\Delta K^{2}Y(t))$. However, the ballistic example emphasises that when the accumulated phase spans a large range, monotonic variations in $Y(t)$ lead to oscillations in the complex factor entering the ISF. The oscillations shown in the ballistic limit translate to a shift of the scattering function in the energy domain.

An oscillatory imaginary signal (polarization) is routinely seen in helium-3 surface spin echo measurements of surface dynamics, the experimental  context that provided the impetus for the present investigation. However, the physical origin of the imaginary oscillations is usually scattering from surface phonons \cite{Jardine2007JPCM,Hedgeland2009PRB}. Additionally, based on the general result (\ref{eqn:recoilFromVACF}) and the VACF for Langevin dynamics in a harmonic well \cite{Vega2004JPCM}, the ISF associated with an isolated underdamped bound adsorbate will exhibit an oscillatory imaginary part. The oscillations described in Figure \ref{fig:quantumRecoil} are related but do not originate from the scattering centre being permanently bound, since we are discussing an unconfined particle. It has been described classically how oscillations in GLE correlation functions can arise from transient wells created by the bath coupling, an effect known as solvent caging \cite{Martens2002JCP}. Therefore, the results in Figure \ref{fig:quantumRecoil} describe how oscillations in the recoil function $Y(t)$ come about for diffusion in a completely flat potential energy landscape, as a result of the finite correlation time in the fluctuating bath degrees of freedom. A confluence of the results in the present work, models for surface phonon lineshapes \cite{MartinezCasado2010JPCM} and continuing experimental refinements for the efficient measurement of imaginary polarization \cite{Tamtogl2018RSI} and complete spectra \cite{Kole2010JPCM,Jones2016SciRep} could allow the experimental disentanglement of the effects in future for suitable systems of low-mass adsorbates. Although the separation of clean surface phonon effects and effects due to adsorbate recoil would be a serious challenge, the concept is at least feasible in helium scattering owing to the large cross section for diffuse scattering from isolated adsorbates \cite{Farias1998RepProgPhys}. The general concept that a coupling between two distinct components of a surface system can be probed via scattering from one component, is interestingly familiar in the context of helium scattering and the electron-phonon interaction \cite{Tamtogl2017PRB}. The potential of detailed scattering measurements to resolve details of the adsorbate/bath coupling is very significant given the ongoing interest in separating out and quantifying the different contributions to atomic-scale dissipation during surface diffusion \cite{Rittmeyer2016PRL}, where memory effects in dissipation are likely to be indicative of the coupling being predominantly to phononic rather than electronic degrees of freedom.

To relate our results to the broader context of non-Markovian systems, we briefly draw attention to an alternative mechanism by which oscillations in the imaginary part of correlation functions appear due to the nature of system-environment coupling. In the model considered throughout the present article, the system co-ordinate is directly coupled to a large number of bath modes. In the context of optical spectra associated with two-state electronic transitions in dye molecules in solution, the physical situation motivates a different family of non-Markovian coupling models. A vibrational solute mode is linearly coupled to the electronic states such that it experiences a net force when the dye is in the excited state. The solute mode is then coupled to a continuum of solvent modes acting as the heat bath \cite{Li1994JACS}, allowing the vibrational coordinate to relax to a new equilibrium displacement in the electronic excited state that reduces the optical energy gap for subsequent photon emission (Stokes shift) \cite{Bosma1990PRA}. The solute mode (special molecular mode) is taken to undergo quantum Brownian motion subject to Langevin friction, and because its motion is directly proportional to the instantaneous optical transition energy, its correlation function $g(t)$ can be measured by spectroscopic means. In broad analogy with the ISF for quantum diffusion, the imaginary part of $g(t)$ leads to spectral shifts that can be resolved at low temperature. If the imaginary part of the correlation function is underdamped (has memory), one may resolve a progression of vibronic sidebands, i.e. see the quantum nature of the environment, but overdamped motion leads to a continuous broadening of the spectral line and a Stokes shift. At very high temperatures, the optical correlation function becomes essentially real-valued (like the classical ISF), and the Stokes shifts can no longer be resolved in optical spectra described by the model above \cite{Bosma1990PRA}.

\section{Conclusions}

Making use of a normal modes transformation, and re-exponentiation result for a single normal mode, we have derived analytical expressions for the intermediate scattering function (ISF) of a quantum particle diffusing in a flat or harmonic potential landscape, linearly interacting with a harmonic bath. The results are presented in the form of an exact relationship between the classical velocity autocorrelation, and the real and imaginary parts of the exponent of the ISF. The results are valid for arbitrary memory friction and therefore extend previous work carried out in the quantum Langevin framework where the imaginary part of the ISF exponent was calculated in a Markovian limit. The results allow a straightforward reference calculation of the quantum ISF for arbitrarily strong and non-Markovian friction, which could be used to benchmark more general but approximate methods for calculating quantum correlation functions. We have provided detailed results in closed form for the special case of unconfined diffusion subject to a memory friction kernel of overall strength $\gamma$, decaying exponentially in time with a rate $\omega_{c}$ that quantifies memory effects. The detailed behaviour of the imaginary part of the exponent of the ISF depends on both $\gamma$ and $\omega_{c}$. However, the long time limit is independent of $\omega_{c}$, and the short-time behaviour is independent of both $\gamma$ and $\omega_{c}$, consistent with universal ballistic behaviour on a short enough time scale.

\begin{acknowledgements}

PT thanks the UK EPSRC for doctoral funding under the award reference 1363145, which enabled the majority of the present work. PT thanks Dr John Ellis for helpful comments on sections of the manuscript, and Prof. Salvador Mir\'{e}t-Artes for highlighting some important prior context.

\end{acknowledgements}

\appendix*
\color{blue}
\section{Normal modes transformation}

We provide, for convenient reference, some brief additional details on the steps leading from the form of the Hamiltonian in Equation \ref{eqn:calLegHam} to Equation \ref{eqn:normalModesHamiltonian}, from the perspective of co-ordinates and momenta. The presentation here is nothing new, but is included for clarity and to explain the transformation in the prevailing framework and notation of the main text. As stressed in the main text, there is no need to actually carry out the transformation described here: to justify the derivation of the central results of the article, it is sufficient that the transformation is legitimate and can be carried out in principle.

The Hamiltonian of Equation \ref{eqn:calLegHam} can be written as a quadratic form over the co-ordinate and momentum operators. Write all the co-ordinates of the global system as a column vector $\mathbf{x}$ in which the first element of the vector is the system co-ordinate $x$, and the rest are the $x_{\alpha}$. Define $\mathbf{p}$ in the analogous way with the corresponding momenta. Then, the Hamiltonian \ref{eqn:calLegHam} can be expressed as
{\begin{equation}
H=\frac{1}{2m}\mathbf{p}^{T}\mathbf{p}+\frac{1}{2}\mathbf{x}^{T}\mathbf{Vx}\, \mathrm{,}
\end{equation}}
where $m$ is still the particle mass, and $\mathbf{V}$ is a real symmetric matrix. We have assumed that the mass of every bath oscillator mode is the same as the particle mass, which according to arguments in the main text leads to no loss of generality.

Given that $\mathbf{V}$ is real and symmetric, there exists an orthogonal matrix $\mathbf{O}$ such that $\mathbf{O}^{T}\mathbf{VO}=\mathbf{D}$ where $\mathbf{D}$ is diagonal. Let $\mathbf{y}$ be a column vector representing a set of operators $y_{k}$, constructed from $\mathbf{x}$ by a linear transformation 
\begin{equation}
\label{eqn:linearCoordinatesTransform}
\mathbf{y}=\mathbf{Ox}
\end{equation}
Define analogously for the momentum operators 
\begin{equation}
\label{eqn:linearMomentaTransform}
\mathbf{q}=\mathbf{Op}
\end{equation}
representing a set of operators $q_{k}$. Then the Hamiltonian can be written as
{\begin{equation}
H=\frac{1}{2m}\mathbf{q}^{T}\mathbf{q}+\frac{1}{2}\mathbf{y}^{T}\mathbf{Dy}\, \mathrm{.}
\end{equation}}
Since $\mathbf{D}$ is diagonal, $H$ is simply the sum of independent oscillator Hamiltonians, as given by Equation \ref{eqn:normalModesHamiltonian}, as long as the collections of operators $\{y_{k}\}$ and $\{q_{k}\}$ satisfy the commutation relations defining them as independent co-ordinate and momentum operators:
\begin{equation}
\label{eqn:zeroCommutator}
[y_{k},y_{l}]=[q_{k},q_{l}]=0 \, \, \textrm{,}
\end{equation}
and
\begin{equation}
\label{eqn:deltaCommutator}
[y_{k},q_{l}]=i\hbar \delta_{k,l}\,\textrm{,}
\end{equation}
where $\delta_{kl}$ is the Kronecker delta symbol. It is straightforward to show that if the original sets of operators represented by $\mathbf{x}$ and $\mathbf{p}$ obeyed the correct commutation relations for independent degrees of freedom, then so do $\{y_{k}\}$ and $\{q_{k}\}$. The relations \ref{eqn:zeroCommutator} are trivially satisfied because linear combinations of commuting operators also commute. The position-momentum commutators (\ref{eqn:deltaCommutator}) can be found by writing out the linear transformations \ref{eqn:linearCoordinatesTransform} and \ref{eqn:linearMomentaTransform} as $y_{k}=O_{k,a}x_{a}$ and $p_{k}=O_{k,a}x_{a}$ assuming the summation convention. Then, the commutators can be worked out as
\begin{equation}
\label{eqn:commutatorWorking}
[y_{k},q_{l}]=O_{k,a}O_{b,l}[x_{a},p_{b}]=i\hbar O_{k,a}O_{b,l}\delta_{a,b}=i\hbar O_{k,a}O_{a,l}\,\textrm{.}
\end{equation}
The defining property of an orthogonal matrix is that $\mathbf{O}\mathbf{O}^{T}=\mathbf{I}$, or $O_{k,a}O_{a,l}=\delta_{k,l}$, and therefore the commutators $[y_{k},q_{l}]$ satisfy the required relation \ref{eqn:deltaCommutator}.


\color{black}

\end{document}